\documentclass[page-classic]{epl2}

\title{Quantum electrodynamic effects on the two-stream instability}
\shorttitle{Title} %Insert here a short version of the title if it exceeds 70 characters

\author{Antoine Bret\inst{1,2}}
\shortauthor{A. Bret}

\institute{
  \inst{1} ETSI Industriales, Universidad de Castilla-La Mancha, 13071 Ciudad Real, Spain\\
  \inst{2} Instituto de Investigaciones Energ\'{e}ticas y Aplicaciones Industriales, Campus Universitario de Ciudad Real,  13071 Ciudad Real, Spain
}

\abstract{
We consider the quantum electrodynamic corrections to the two-stream instability. We find these corrections vanish at first order unless a guiding magnetic field $\mathbf{B}_0$ is considered. With respect to the classical version of the instability, quantum electrodynamic effects reduce the most unstable wave vector and its growth rate by a factor $\sqrt{1+\xi}$, with $\xi = \frac{\alpha}{9\pi} (B_0/B_{cr})^2$, where $\alpha$ is the fine-structure constant and $B_{cr}$ the Schwinger critical magnetic field. Although derived for a cold system, these results are valid for the kinetic case. The results are valid in the range $\xi \ll 1$ and, actually, up to linear corrections in $\xi$.
}

\begin{document}

\maketitle

The rise of extreme plasma physics \cite{DiPiazzaRMP2012,Uzdensky2014,SilvaEPS2021}, where fields are so intense that quantum electrodynamic (QED) effects start to play a role, suggests we should revisit some basic processes in plasma physics accounting for these effects \cite{MarklundEPL2005,LundinPoP2007,DiPiazzaPoP2007,HuPoP2012,ChenPoP2013}.

Among these basic processes stand the two-stream instability (TSI) which relates to the instability of 2 counter-streaming beams. Streaming instabilities have been considered in the context of pulsar emissions \cite{Gedalin2002,AsseoPPCF2003}, where fields can reach $10^{12}$ Gauss, close to the critical Schwinger magnetic field $B_{cr}= m_e^2  c^3 / q \hbar = 4.4 \times 10^{13}$ Gauss or $4.4 \times 10^9$ Tesla  \cite{Schwinger1951}. Other astrophysical settings, like neutron stars, involve fields that can reach $10^{14}$ Gauss, where corrections of even higher orders than those explained here would be required \cite{DongSSRv2015}.

\begin{figure}
\onefigure{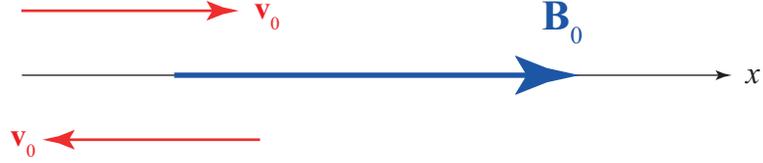}
\caption{System considered (1D): two symmetric, cold, relativistic electron beams at $\pm \mathbf{v}_0$ over a neutralizing background of fixed ions. A guiding magnetic field $\mathbf{B}_0$ is included which eventually yields a non-vanishing correction to the TSI.}
\label{fig.setup}
\end{figure}

The goal of this letter is to assess QED effects in relation to this instability. The system studied is sketched on Figure \ref{fig.setup}. We consider two symmetric, cold, relativistic electron beams over a neutralizing background of fixed ions. The calculations are conducted in the reference frame of the ions. In this frame, the electrons have density $n_0$. They initially stream at $\pm \mathbf{v}_0$ with Lorentz factor $\gamma_0=(1-v_0^2/c^2)^{-1/2}$.

A guiding magnetic field $\mathbf{B}_0 = B_0 \mathbf{x}$ is included in the analysis. For a classical system, such a field has no effect on the TSI because TSI has particles oscillate along the field, hence cancelling  the Lorentz force. Here, as shall be checked in the sequel, QED effects arise from $\mathbf{B}_0$ and do alter the TSI. At first order these effects vanish for $B_0=0$.

We first briefly remind the derivation of the classical TSI in order to clearly see where QED corrections come into play.

For the 1D system pictured on Figure \ref{fig.setup}, the dispersion equation for the classical TSI can be derived from a two-fluids formalism and the Poisson equation. One writes for each species ``$i$'',
\begin{eqnarray}
\frac{\partial n_i}{\partial t} + \nabla \cdot ( n_i \mathbf{v}_i) &=& 0, \label{eq:conser}  \\
\frac{\partial \mathbf{p}_i}{\partial t} + (\mathbf{v}_i\cdot \nabla) \mathbf{p}_i &=& q_i\left( \mathbf{E} + \frac{\mathbf{v}_i \times \mathbf{B}}{c}  \right), \label{eq:euler}
\end{eqnarray}
with,
\begin{equation}\label{eq:poissonclass}
\nabla \cdot  \mathbf{E}  = 4\pi \rho_c,
\end{equation}
where $\rho_c=\sum_i q_i n_i$ is the classical charge density. To analyze  the TSI, these equations are perturbed in the direction of the flow. Eqs. (\ref{eq:conser}-\ref{eq:poissonclass}) are therefore linearized assuming small perturbations $\propto \exp (ikx-i\omega t)$.  Such perturbations could be seeded by a slight variation of the density along the path of the beam, or even by the spontaneous fluctuations of the plasma \cite{BretPoP2013}. Eqs. (\ref{eq:conser},\ref{eq:euler}) give the first order density and velocity perturbations,
\begin{eqnarray}
n_{1\pm} &=& n_0\frac{k v_{1\pm}}{\omega \pm kv_0}, \nonumber\\
v_{1\pm} &=& \frac{q}{m_e\gamma_0^3(\omega \pm kv_0)} i  E_1,
\end{eqnarray}
where $m_e$ is the electron mass and subscripts ``1'' refer to perturbed quantities. Altogether these two equations give for $n_{1\pm}$,
\begin{equation}
n_{1\pm} = n_0\frac{q}{m_e\gamma_0^3(\omega \pm kv_0)^2} i k E_1,
\end{equation}
which, inserted into the linearized Poisson equation,
\begin{equation}\label{eq:PoiClass}
ikE_1 = 4\pi q n_{1+} +  4\pi q n_{1-},
\end{equation}
gives the dispersion equation for the classical TSI (see for example \cite{TB2017}, p. 1065 or  \cite{boyd}, p. 239),
\begin{equation}\label{eq:DisperClass}
1 = \frac{\omega_p^2}{\gamma_0^3(\omega-kv_0)^2}   + \frac{\omega_p^2}{\gamma_0^3(\omega+kv_0)^2} ,
\end{equation}
where $\omega_p^2=4\pi n_0q^2/m_e$.

How do QED effects modify this picture?

QED corrections to Eqs. (\ref{eq:conser}-\ref{eq:poissonclass}) only enter the Poisson equation (see Eqs. (1-8) of Ref. \cite{DiPiazzaPoP2007}). Its modified version reads,
\begin{equation}\label{eq:poissonqed}
\nabla \cdot  \mathbf{E} = 4\pi(\rho_c + \rho_{vac}),
\end{equation}
where $\rho_{vac}$ is the vacuum polarization  due to QED effects. It is given by \cite{DiPiazzaRMP2012},
\begin{equation}\label{eq:rhovac}
\rho_{vac} = -\frac{1}{180\pi^2}\frac{\alpha}{c^2B_{cr}^2} \nabla \cdot \left[  2(E^2-c^2B^2)\mathbf{E} +  7c^2(\mathbf{E} \cdot \mathbf{B}) \mathbf{B}  \right],
\end{equation}
where $\alpha$ is the fine-structure constant. The modified Poisson equation (\ref{eq:poissonqed}) can be obtained varying the Lagrangian of matter plus that of the field, where the field' one comes from the Euler-Heisenberg effective Lagrangian. This effective Lagrangian accounts for vacuum polarization effects in the small field limit (\cite{DiPiazzaRMP2012} or \cite{weinberg2005quantum}, p. 32). Consequently, the expression (\ref{eq:rhovac}) assumes $B_0 \ll B_{cr}$.

To establish the QED modified dispersion equation we need to linearize the modified Poisson equation (\ref{eq:poissonqed}), hence linearize $\rho_{vac}$. We therefore write $\mathbf{E}=\mathbf{E}_1$ and $\mathbf{B}=\mathbf{B}_0+\mathbf{B}_1$ where subscript ``1'' pertains to first order perturbations of the quantities, and obtain\footnote{We set $c=1$ in this equation only, to shorten notations.}
\begin{eqnarray}
2(E^2-B^2)\mathbf{E} &+&  7(\mathbf{E} \cdot \mathbf{B}) \mathbf{B} =
                                       2 \left[ E_1^2 \mathbf{E}_1- B_0^2 \mathbf{E}_1  -  B_1^2 \mathbf{E}_1 -  (\mathbf{B}_0 \cdot \mathbf{B}_1) \mathbf{E}_1\right] \\
   &+& 7 \left[  (\mathbf{E}_1 \cdot \mathbf{B}_0)\mathbf{B}_0 +(\mathbf{E}_1 \cdot \mathbf{B}_0)\mathbf{B}_1 + (\mathbf{E}_1 \cdot \mathbf{B}_1)\mathbf{B}_0 + (\mathbf{E}_1 \cdot \mathbf{B}_0)\mathbf{B}_1  \right]. \nonumber
\end{eqnarray}
At first order, the only remaining terms on the right-hand-side are,
\begin{equation}\label{eq:1storder}
2(E^2-c^2B^2)\mathbf{E} +  7c^2(\mathbf{E} \cdot \mathbf{B}) \mathbf{B} = -2c^2B_0^2 \mathbf{E}_1 + 7 c^2(\mathbf{E}_1 \cdot \mathbf{B}_0)\mathbf{B}_0 + \ldots
\end{equation}
For a 1D geometry like that of Figure \ref{fig.setup}, the first order correction thus reads,
\begin{equation}
-2B_0^2 \mathbf{E}_1+7 B_0^2E_{1x}\mathbf{x}  = 5 c^2B_0^2 \mathbf{E}_1.
\end{equation}
Therefore, at first order and for the present 1D case,
\begin{equation}\label{eq:rhovacL}
\rho_{vac} = -\frac{1}{180\pi^2}\frac{\alpha}{c^2B_{cr}^2} \nabla \cdot \left[  5 c^2 B_0^2 \mathbf{E}_1  \right] =
    -   \frac{\alpha}{36\pi^2}\frac{B_0^2}{B_{cr}^2}  \frac{\partial E_1}{\partial x}.
\end{equation}
Noteworthily,  first order corrections to the Poisson equations are zero unless $B_0 \neq 0$.

We can now establish the QED-modified dispersion equation for the TSI. The QED correction to the classical Poisson equation (\ref{eq:PoiClass}) is straightforward and reads,
\begin{equation}\label{eq:PoiQED}
ikE_1 = 4\pi q n_{1+} + 4\pi q n_{1-} - \underbrace{4\pi\frac{\alpha}{36\pi^2}\frac{B_0^2}{B_{cr}^2}}_{\equiv \xi}~ik E_1.
\end{equation}
The strength of the QED corrections is therefore measured by,
\begin{equation}\label{eq:xi}
\xi = \frac{\alpha}{9\pi}\frac{B_0^2}{B_{cr}^2}.
\end{equation}

Using then the same procedure than for the classical case, we obtain the QED counterpart of Eq. (\ref{eq:DisperClass}),
\begin{equation}\label{eq:DisperQED}
1  + \xi = \frac{\omega_p^2}{\gamma_0^3(\omega-kv_0)^2} + \frac{\omega_p^2}{\gamma_0^3(\omega+kv_0)^2} .
\end{equation}
Setting $x=\omega/\omega_p$ and $Z=kv_0/\omega_p$ this equation reads,
\begin{equation}
1 +   \xi  =  \frac{1}{\gamma_0^3(x  - Z )^2} +\frac{1}{\gamma_0^3(x  + Z )^2} .
\end{equation}

\begin{figure}
\onefigure{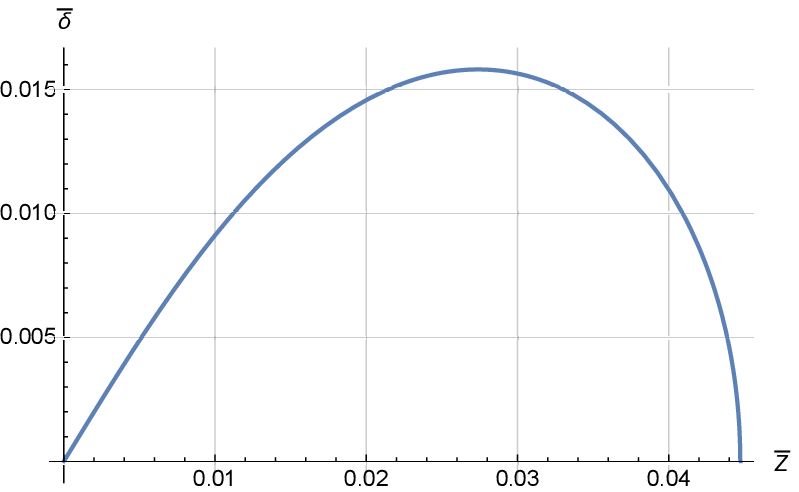}
\caption{Growth rate $\overline{\delta}(\overline{Z})$ from Eq. (\ref{eq:gr}) for $\gamma_0=10$.}
\label{fig.gr}
\end{figure}

The QED corrections eventually amount to substitute $1 \rightarrow 1+\xi$ in the left-hand-side of the classical TSI dispersion equation. Dividing both sides by $1 +  \xi$ now gives
\begin{equation}
1 = \frac{1}{\gamma_0^3(x\sqrt{1 + \xi}  - Z\sqrt{1 + \xi} )^2}  + \frac{1}{\gamma_0^3(x\sqrt{1 + \xi}  + Z\sqrt{1 + \xi} )^2}.
\end{equation}
Therefore, if we set,
\begin{eqnarray}
  \overline{x} &=& x\sqrt{1+ \xi}, \\
  \overline{Z} &=& Z\sqrt{1+ \xi},
\end{eqnarray}
the classical TSI dispersion equation is formally recovered,
\begin{equation}
1 = \frac{1}{\gamma_0^3(\overline{x}  - \overline{Z} )^2} +  \frac{1}{\gamma_0^3(\overline{x}  + \overline{Z} )^2}.
\end{equation}
This equation can be solved exactly for $x$ and gives an unstable mode with growth rate $\overline{\delta}$ plotted on Figure \ref{fig.gr},
\begin{equation}\label{eq:gr}
\overline{\delta}(\overline{Z}) = \sqrt{- \frac{1+\gamma_0^3 \overline{Z}^2-\sqrt{4 \gamma_0^3 \overline{Z}^2+1}}{\gamma_0^3}}.
\end{equation}
This growth rate reaches 0 for $\overline{Z}_{\overline{\delta}=0} = \sqrt{2}/\gamma_0^{3/2}$. It is maximum for $\overline{Z}_m=\sqrt{3}/2\gamma_0^{3/2}$ with $\overline{\delta}(\overline{Z}_m)\equiv \overline{\delta}_m =1/2\gamma_0^{3/2}$. Coming back to the dimensional  variables we eventually find,
\begin{eqnarray}\label{eq:results}
  k_{\delta=0}        &=& \frac{\sqrt{2}}{\gamma_0^{3/2}} ~ \frac{\omega_p/\sqrt{1+ \xi}}{v_0}, \\
  k_m      &=&  \frac{\sqrt{3}}{2\gamma_0^{3/2}} ~ \frac{\omega_p/\sqrt{1+ \xi}}{v_0}, \nonumber \\
  \delta_m &=&\frac{1}{2\gamma_0^{3/2}} ~ \frac{\omega_p}{\sqrt{1+ \xi}}. \nonumber
\end{eqnarray}

In conclusion, QED corrections to the TSI require the presence of a guiding magnetic field to be non-zero at first order. This stands in contrast with the classical TSI where such a field has no effect. From Eqs. (\ref{eq:results}) we see that the most unstable $k$ and the growth rate are both reduced by a factor $\sqrt{1+\xi}$, where $\xi \propto \alpha (B_0/B_{cr})^2$ is given by Eq. (\ref{eq:xi}).

To my knowledge, there is still no deep physical understanding of the TSI, a ``Fermi-like picture'', beyond the mathematical resolution of its dispersion equation. It is then difficult to provide a full physical interpretation of the QED effects on TSI.

What can definitely be said is that in the linear regime, the perturbed field $\mathbf{B}_1$ grows from 0. Therefore, for $\mathbf{B}_0+\mathbf{B}_1$  to approach the Schwinger limit, a finite $B_0$ is required, already close to $B_{cr}$. This is why QED effects for the instability vanish for $B_0=0$. If $B_0=0$, the field $B_1$ alone will not reach $B_{cr}$ in the linear regime of the instability. It would be interesting to determine in future works the conditions required for QED effects to play a role even if $B_0=0$.

The reduction of the growth rate could come from the shielding of charges by vacuum particles. Bare charges attract vacuum particles of the same charge and repel those of opposite charge, so that they are screened (\cite{weinberg2005quantum}, p. 482). Indeed Eqs. (\ref{eq:results}) can be interpreted as a renormalization of the plasma frequency $\omega_p \rightarrow \omega_p /\sqrt{1+ \xi}$, which in turn amounts to a renormalization of the elementary charge $q \rightarrow q /\sqrt{1+ \xi}$, since $\omega_p  \propto q$.

The present result can be straightforwardly extended to the case of electron beams of different densities as Eq. (\ref{eq:PoiClass},\ref{eq:PoiQED}) read the same in that case. Also,  Eqs. (\ref{eq:DisperClass},\ref{eq:DisperQED}) allow to generalize the present results to the kinetic case, since the kinetic dispersion equation can be obtained from Eq. (\ref{eq:DisperQED}) by replacing the sum over the beams on the right-hand-side by an integral in velocity space \cite{Dawson1960}.

The general case where any orientations of $\mathbf{k}$ are considered is more involved because Eq. (\ref{eq:1storder}) will bring more complex corrections to the dielectric tensor.

\acknowledgments
A.B. acknowledges support by grants ENE2016-75703-R from the Spanish
Ministerio de Ciencia, Innovaci\'{o}n y Universidades and SBPLY/17/180501/000264 from the Junta de
Comunidades de Castilla-La Mancha. Thanks are due to Antonino DiPiazza for valuable inputs.

%\bibliographystyle{eplbib}
%\bibliography{BibBret}

\end{document}